\documentclass[onecolumn, 10pt,english,aps]{revtex4}
\usepackage[T1]{fontenc}
\usepackage{color}
\usepackage{array}
\usepackage{graphicx}
\usepackage[all,dvips]{xy}
\usepackage{amssymb}

\usepackage{babel}

\begin{document}

\title{Perpetual points: New tool for localization of co--existing attractors in dynamical systems}

\author{Dawid Dudkowski$^1$, Awadhesh Prasad$^2$, and Tomasz Kapitaniak$^1$}
\affiliation{$^{1}$Division of Dynamics, Technical University of Lodz, Stefanowskiego 1/15, 90--924 Lodz, Poland}
\affiliation{$^{2}$Department of Physics and Astrophysics, University of Delhi, Delhi 110007, India}

\begin{abstract}
Perpetual points (PPs) are special critical points for which the magnitude of acceleration describing dynamics drops to zero, while the motion is still possible (stationary points are excluded), e.g. considering the motion of the particle in the potential field, at perpetual point it has zero acceleration and non--zero velocity. We show that using PPs we can trace all the stable fixed points in the system, and that the structure of trajectories leading from former points to stable equilibria may be similar to orbits obtained from unstable stationary points. Moreover, we argue that the concept of perpetual points may be useful in tracing unexpected attractors (hidden or rare attractors with small basins of attraction). We show potential applicability of this approach by analysing several representative systems of physical significance, including the damped oscillator, pendula and the Henon map. We suggest that perpetual points may be a useful tool for localization of co--existing attractors in dynamical systems.
\vspace{\baselineskip}

\end{abstract}

\maketitle

\section{Introduction}
In 2015, Prasad [Prasad, 2015] has introduced the concept of perpetual points (PPs), which are a new type of critical points in dynamical systems. They have been defined as the points in phase space, in which the acceleration of the system becomes zero, while velocity remains non--zero. The simpliest representation of such state can be imagined as the motion of a particle in the potential field. At perpetual point the particle has zero acceleration, but its motion is still possible (non--zero velocity). In the original work, many useful applications of PPs have been suggested.

In [Prasad, 2016] the topological properties of considered points are studied. The analysis of presence and co--existence of perpetual points in single and coupled van der Pol--Duffing oscillators can be found in [Dudkowski et al., 2015]. Some of the limitations related to the application of proposed concept have been discussed by Jafari et al. [2015] and Nazarimehr et al. [2017], while in [Ueta et al., 2015] authors suggest a connection between slow perpetual points and a phenomenon known as the attractor ruin.

From the very beginning, it has been stated that perpetual points may be useful in localization of co--existing attractors. Such property of considered objects may be particularly relevant in the systems with hidden oscillations [Dudkowski et al., 2016; Chaudhuri \& Prasad, 2014; Leonov et al., 2011; Leonov \& Kuznetsov, 2013; Wei et al., 2014; Wei \& Zhang, 2014; Wei et al., 2015; Wei et al., 2015b; Wei et al., 2016; Wei et al., 2016a]. Multistability of states have been recently widely observed in many dynamical models and thoroughly studied by the researchers [Pisarchik \& Feudel, 2014].

The initial results of our research work on perpetual points have been presented in [Dudkowski et al., 2016], where we studied the potential usage of those points in localization of co--existing attractors. We have focused especially on the unexpected attractors, like hidden states or rare attractors, which basins of attraction are very small. In this paper, we present the extensive analysis of this property of perpetual points. We study several representative models of physical meaning, including damped oscillators and pendula. We also analyze more artificial systems like chaotic flows or Henon map. Models of this type are the basis of studies of dynamical systems and results obtained for them may allow us to better understand the concept of perpetual points and reasons for detection of co--existing states.
\vspace{\baselineskip}

\begin{figure}
\includegraphics[scale=0.5]{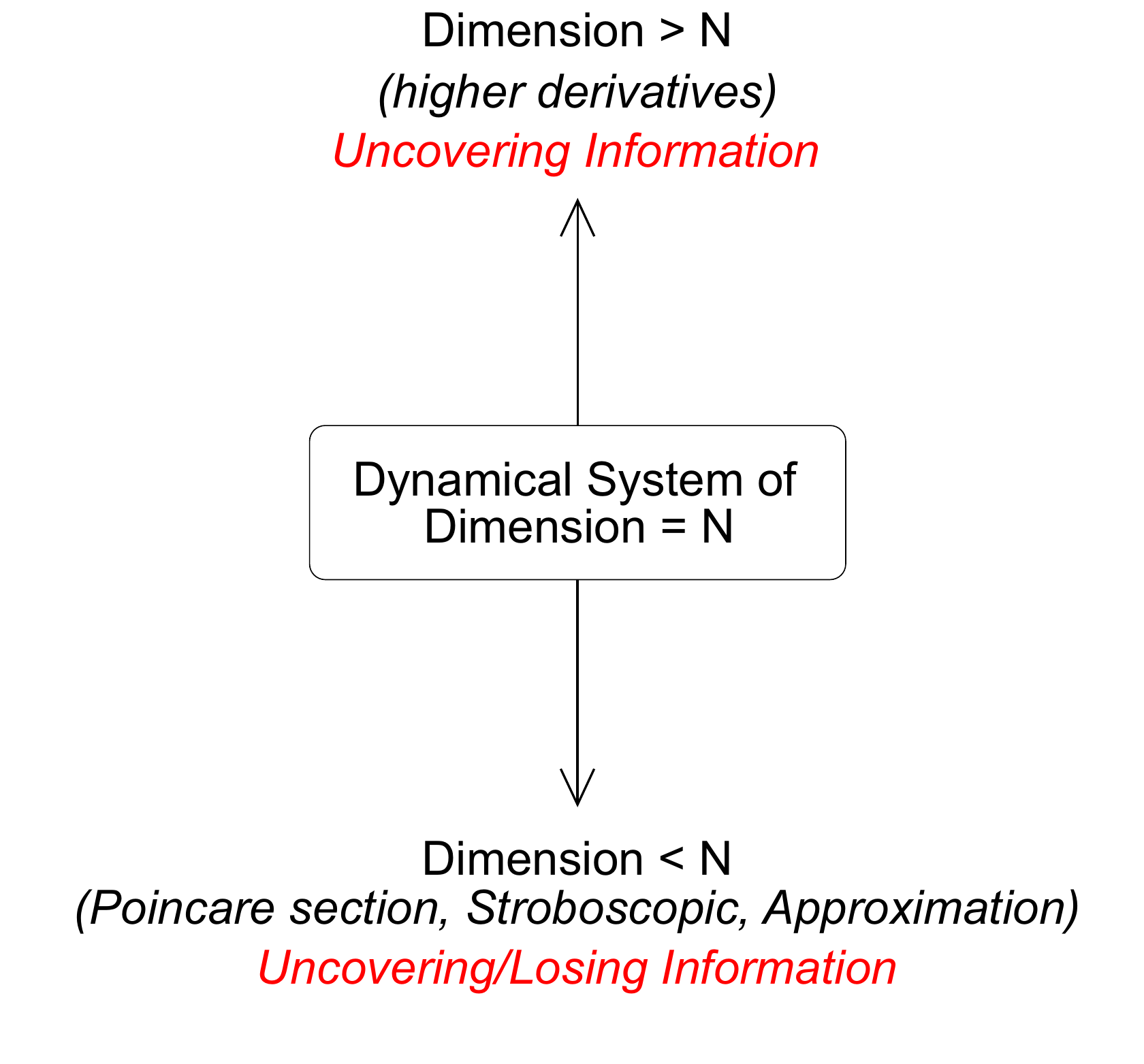}
\caption{(color online). Two methods of extracting information about the system dynamics: decrease of dimension (lower arrow) and increase of dimension (upper arrow).}
\label{fig0}
\end{figure}

In Fig.~\ref{fig0} one can observe two possible ways of extracting information about the system dynamics. The first group, marked by the lower arrow is based on lowering the dimension of considered model. This procedure is especially used in high--dimensional problems, when model is complex and it is very difficult to analyze the behavior of the full system. A typical method of this class is the Poincare section, when the trajectory points are projected on a subspace of phase space. Stroboscopic maps and approximation methods are also widely used. Although some of the information about the system can be uncovered, a natural issue related with this kind of procedures is that if we decrease the dimension, some of the information about the system can be also lost. Because the model is simplified, it may influence on performed research and disrupt final conclusions.

Our motivation to introduce perpetual points is to describe a method that can uncover some information about the system, without simplifying it and losing what is already known. As opposed to decrease of dimension, this can be achieved by increase of phase space (upper arrow in Fig.~\ref{fig0}). The most natural way is to consider higher derivatives of considered system, as in the case of perpetual points when second derivative is studied. The original $N$--dimensional system is considered in additional $N$ dimensions, which allows us to uncover new information on considered problem without perturbing it by approximations. Moreover, there is no need to include new conditions, because initial values for additional dimensions can be obtained from the original ones, which are already known. Our aim in this paper is to understand the original system when considering additional $N$ dimensions.

\section{Results}
Firstly, we have investigated simple one--dimensional systems. It should be emphasized that in this paper the dimension of the system will be understood as the number of variables, which are needed to represent the potential function of particular model.

Such systems can be represented generally by the family of unforced damped oscillators, i.e. equations of the form:
\begin{equation}
\ddot z + \alpha \dot z + \frac{d}{dz} V(z)  = 0,
\label{potGeneral}
\end{equation}
where $V$ is the potential and $\alpha$ is the damping coefficient. In [Dudkowski et al., 2016] we have shown that if a system (\ref{potGeneral}) has a perpetual point, then the value of the first coordinate of such a point (points are considered in $(z, \dot z)$ phase space) is located exactly at the inflection of the potential energy function $V$.

Due to the fact that the trajectories starting from inflections converge to the stable solutions through the shortest paths (decreasing the magnitude of potential energy in the system), the inflection and stationary points are naturally connected. According to this, the relation between inflection and perpetual points may uncover the potential usage of the latter ones in localization of co--existing attractors.

We have considered two systems of the form (\ref{potGeneral}). The first one, a plane pendulum [Blekhman, 2003; Barger \& Olsson, 1973; Kibble \& Berkshire, 2004; Baker \& Blackburn, 2005] is one of the most elementary models in dynamical systems theory and can be described by the following equation:
\begin{equation}
m l^2 \ddot{\varphi} + m g l \sin \varphi + d \dot{\varphi} = 0,
\label{pendulum1}
\end{equation}
where $m, l, d$ are mass, length and damping coefficient respectively of the pendulum, while $g$ is the standard acceleration due to gravity. We fix $m=1$, $l=1$, $d=0.2$ and $g=1$ (value of $g$ has been chosen for the better clarity of Figures). Note that the angular position $\varphi$ of the pendulum has been considered in interval $\varphi \in [0, 2 \pi)$. 

The second one--dimensional system that we consider is the Duffing oscillator [Ott, 1993; Kovacic \& Brennan, 2011; Parlitz \& Lauterborn, 1985; Sabarathinam et al., 2013], which dynamics is given by:
\begin{equation}
\ddot{x} + b \dot{x} - a x + x^3 = 0,
\label{Duff}
\end{equation}
where $a$ and $b$ are the parameters, fixed at $a=2, b=1$.

\begin{figure}
\includegraphics[scale=0.55]{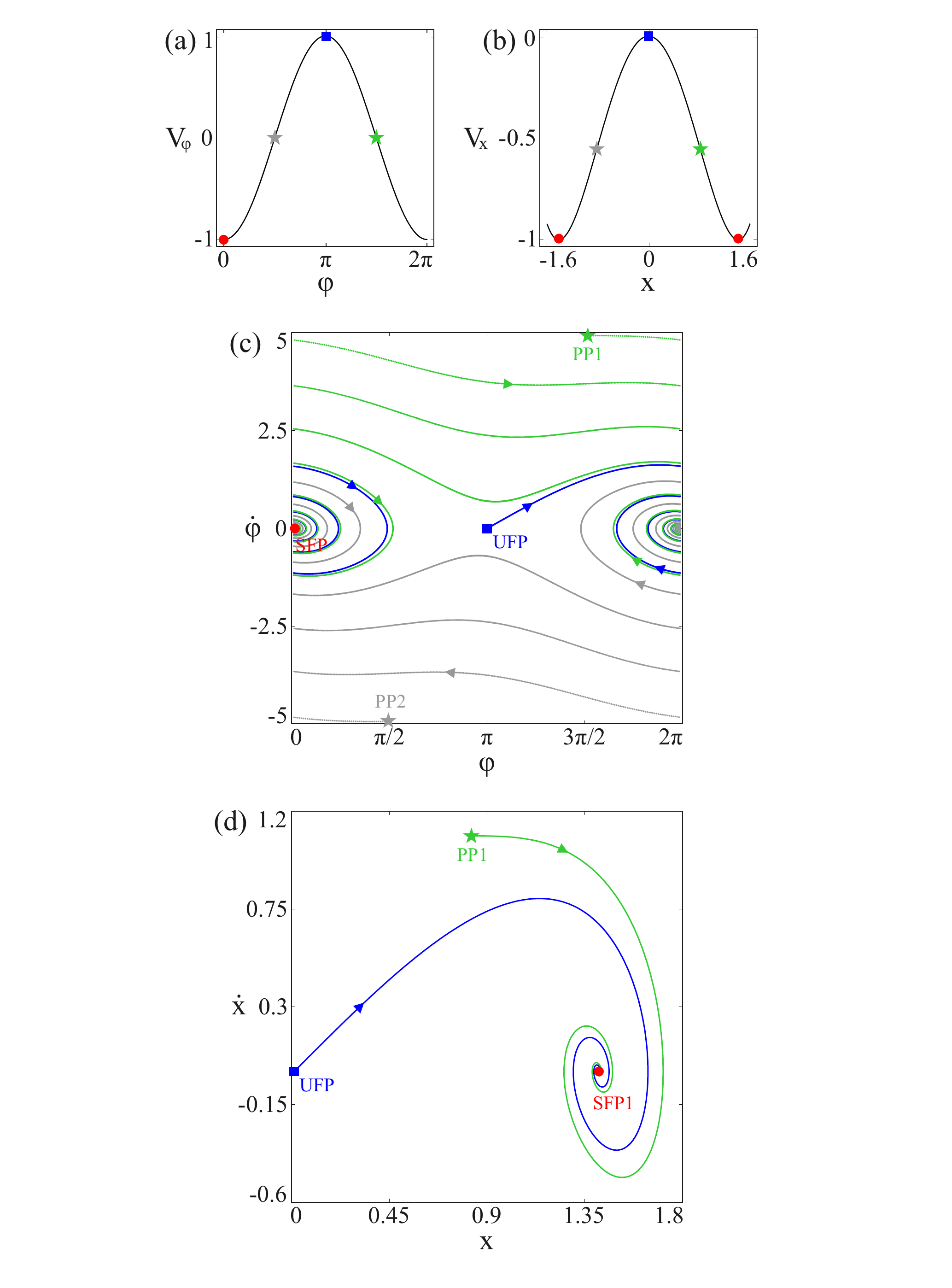}
\caption{(color online). In (a--b) the potentials of systems (\ref{pendulum1}) and (\ref{Duff}) are shown in left and right panel respectively. Stable/unstable fixed points are presented as red circles/blue squares, while perpetual points are denoted by green and gray stars. Trajectories starting from critical points of pendulum (\ref{pendulum1}) and the Duffing oscillator (\ref{Duff}) are shown as curves in (c) and (d) respectively. Color code is the same as in (a--b), and the arrows denote the direction of motion.}
\label{fig1}
\end{figure}

Results of our investigations on systems (\ref{pendulum1}, \ref{Duff}) are presented in Fig.~\ref{fig1}.
In Fig.~\ref{fig1}(a--b) the potentials of (\ref{pendulum1}) and (\ref{Duff}) are shown in left and right panel respectively. For the pendulum, the potential energy is described by $V_{\varphi} = - m g l \cos \varphi$, while the potential of the Duffing oscillator is given by $V_x = 1/4 x^4 - 1/2 a x^2$. The red circles/blue squares correspond to stable/unstable equilibria of the systems, while the green and gray stars denote perpetual points. As one can see, stationary points are naturally in the extrema of presented functions, while perpetual points are at inflections. In the case of the pendulum, both points $(\pi / 2, - m g l / d)$ and $(3 \pi / 2, m g l / d)$ lead to the stable $(\varphi, \dot{\varphi}) = (0,0)$ equilibrium (due to considered interval of $\varphi$ values, state $\varphi = 2 \pi$ is identified as $\varphi = 0$). On the other hand, the Duffing oscillator stable fixed points $(-\sqrt{a},0)$ and $(\sqrt{a},0)$ can be located using perpetual points $(-\sqrt{\frac{a}{3}}, -\frac{2a}{3b} \sqrt{\frac{a}{3}})$ and $(\sqrt{\frac{a}{3}}, \frac{2a}{3b} \sqrt{\frac{a}{3}})$ respectively.

We have studied closely the trajectory paths when systems run from different types of critical points. In Fig.~\ref{fig1}(c) one can observe the phase portrait of pendulum (\ref{pendulum1}). Orbits starting from unstable fixed point UFP=$(3.1416, 0)$ is shown as blue curve, and converges to the stable equilibrium SFP=$(0, 0)$ (red). On the other hand, trajectories that begin from perpetual points PP1=$(4.7124, 5)$ and PP2=$(1.5708, -5)$ are presented in green and gray respectively (color code corresponds to Fig.~\ref{fig1}(a)). The arrows in Fig.~\ref{fig1}(c) denote the direction of motion. As one can see, although both perpetual points lead to the same state, the curve starting from PP1 additionally converge to the path denoted by UFP. This can be observed especially near the SFP. When pendulum oscillates around the equilibrium state, both trajectories (green and blue) get very close to each other and the motion is almost the same. Similarly, in Fig.~\ref{fig1}(d) phase dynamics of the Duffing oscillator (\ref{Duff}) for $x>0$ is presented (symmetrical portraits can be obtained for $x<0$). The unstable fixed point UFP=$(0, 0)$ and perpetual point PP1=$(0.8165, 1.0887)$ (blue square and green star respectively) both lead to the stable state SFP=$(1.4142, 0)$ (red circle). The paths of the orbits (green and blue curves) are similar, but the difference around SFP is not as small as in pendulum case.
\vspace{\baselineskip}

Perpetual points can be found not only in the simplest dynamical systems, like the ones presented above, but also in more complex models. In our research we have considered two--dimensional systems, i.e. the ones for which two variables are needed to describe the potential.
An example of such a case is the double pendulum [Blekhman, 2003; Barger \& Olsson, 1973; Kibble \& Berkshire, 2004; Baker \& Blackburn, 2005]. Its dynamics is given by following formulas (obtained from Lagrange's equations of motion):
\begin{eqnarray}
   \left\{
     \begin{array}{l}
     (m_1+m_2) l^{2}_{1} {\ddot{\varphi}}_1 + m_2 l_1 l_2 \cos (\varphi_1 - \varphi_2) {\ddot{\varphi}}_2 + m_2 l_1 l_2 \sin (\varphi_1 - \varphi_2) {\dot{\varphi}}^{2}_{2} + (m_1+m_2) g l_1 \sin \varphi_1 + d_1 {\dot{\varphi}}_1 + d_2 ({\dot{\varphi}}_1 - {\dot{\varphi}}_2) = 0,\\
		 m_2 l^{2}_{2} {\ddot{\varphi}}_2 + m_2 l_1 l_2 \cos (\varphi_1 - \varphi_2) {\ddot{\varphi}}_1 - m_2 l_1 l_2 \sin (\varphi_1 - \varphi_2) {\dot{\varphi}}^{2}_{1} + m_2 g l_2 \sin \varphi_2 - d_2 ({\dot{\varphi}}_1 - {\dot{\varphi}}_2) = 0,
     \end{array}
   \right.
	\label{pendulum2}
\end{eqnarray}
where $\varphi_1$ is the angular position of upper pendulum (1st) suspended to the beam, while $\varphi_2$ describes the angular position of lower pendulum (2nd) coupled with the upper one. Parameters $m_1 = m_2 = 1$, $l_1 = l_2 = 1$ and $d_1 = d_2 = 0.2$ denote the masses, lengths and damping coefficients of 1st and 2nd pendulum respecively, while $g=1$ is the standard acceleration.

\begin{figure}
\includegraphics[scale=0.55]{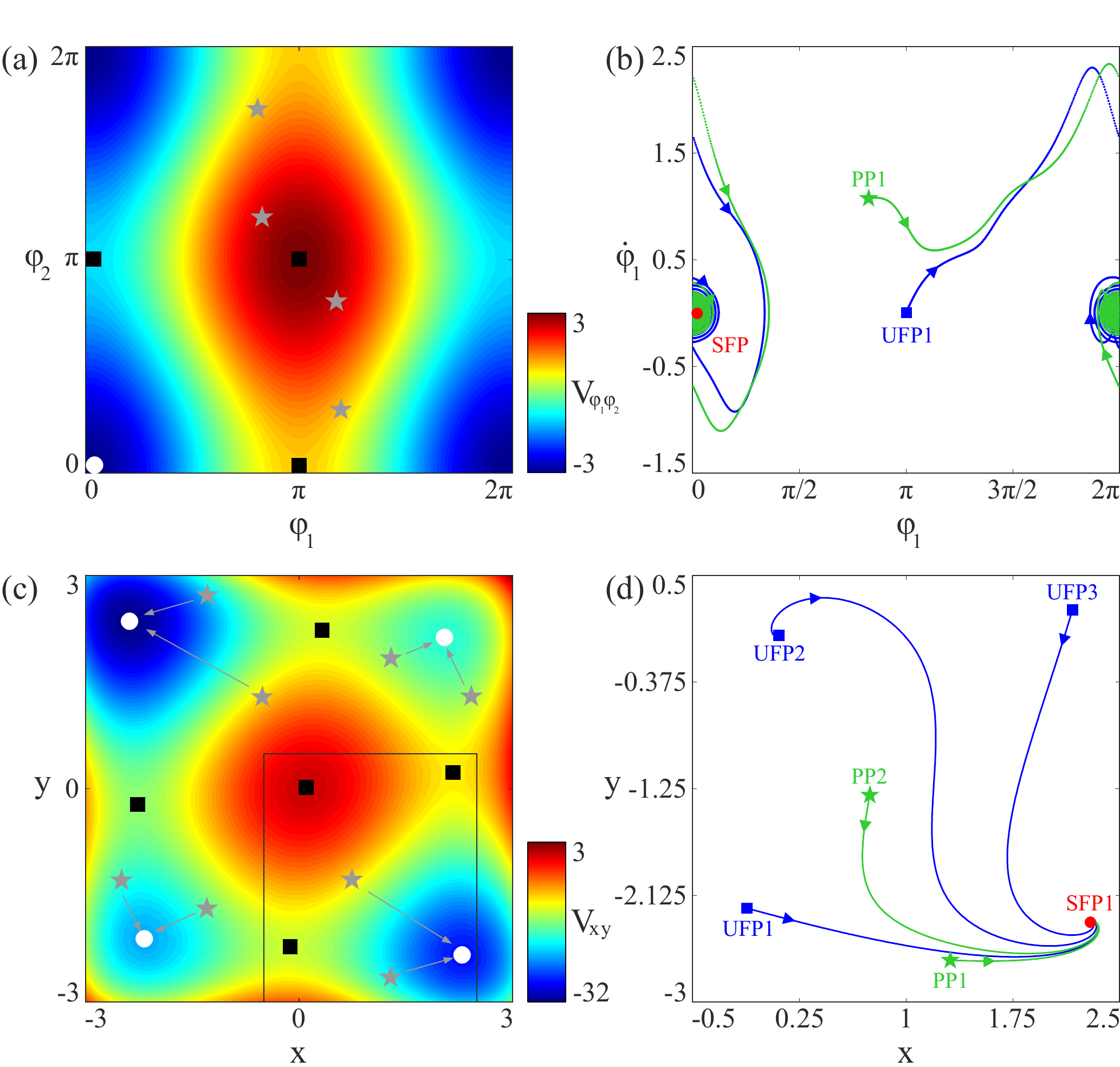}
\caption{(color online). The color plots of potential functions of systems (\ref{pendulum2}) and (\ref{quasiSys}) are shown in (a) and (c) respectively. White circles denote stable fixed points, while black squares correspond to the unstable ones. Perpetual points are marked as gray stars. On the other hand, trajectories beginning from some of the critical points of both systems are presented in (b) (double pendulum) and (d) (polynomial system). Here, stable/unstable equilibria are shown as red circles/blue squares, while perpetual points are presented as green stars.}
\label{fig2}
\end{figure}
 
In Fig.~\ref{fig2}(a--b) one can observe the results obtained from the study of system (\ref{pendulum2}). The potential energy in this case is given by $V_{\varphi_1 \varphi_2} = - m_1 g l_1 \cos \varphi_1 - m_2 g (l_1 \cos \varphi_1 + l_2 \cos \varphi_2)$ and has been shown as color plot in Fig.~\ref{fig2}(a). Color scale describes the values of potential for $(\varphi_1, \varphi_2) \in {[0, 2 \pi)}^2$, while markers presented in grayscale denote projection of critical points of the double pendulum on $(\varphi_1, \varphi_2)$ plane. Stable equilibrium $(\varphi_1, {\dot{\varphi}}_1, \varphi_2, {\dot{\varphi}}_2) = (0,0,0,0)$ shown as white circle co--exists with the unstable fixed points $(\pi,0,0,0)$, $(0,0,\pi,0)$ and $(\pi,0,\pi,0)$, shown as black squares. We have found four perpetual points in system (\ref{pendulum2}), i.e. $(2.5922,1.0792,3.7593,-1.3803)$, $(3.691,-1.0792,2.5239,1.3803)$, $(2.5262,1.4873,5.3539,2.0784)$, and $(3.757,-1.4873,0.9293,-2.0784)$, which are marked by gray stars. As can be seen, obtained points exhibit similar symmetry properties as in the case of single pendulum (\ref{pendulum1}). Namely, if $(\varphi^{PP}_{1}, {\dot{\varphi}}^{PP}_{1}, \varphi^{PP}_{2}, {\dot{\varphi}}^{PP}_{2})$ is a perpetual point, then point $(2 \pi - \varphi^{PP}_{1}, -{\dot{\varphi}}^{PP}_{1}, 2 \pi - \varphi^{PP}_{2}, -{\dot{\varphi}}^{PP}_{2})$ is perpetual also. In Fig.~\ref{fig2}(b) we have compared the trajectories beginning from unstable equilibrium UFP1=$(3.1416, 0, 3.1416, 0)$ (blue square) and perpetual point PP1=$(2.5922,1.0792,3.7593,-1.3803)$ (green star), both leading to the stable state SFP=$(0, 0, 0, 0)$ (red circle). Results are shown for the upper pendulum (projection of phase space on $(\varphi_{1}, {\dot{\varphi}}_{1})$ plane) and similarly to Fig.~\ref{fig1}, arrows denote the direction of motion. As can be seen, the scenario is similar to the one observed for the single pendulum (\ref{pendulum1}). Orbit obtained from PP1 (green curve) gets close to the trajectory denoted by UFP1 (blue one), and their shape is comparable. The behavior of system (\ref{pendulum2}) when it starts from both kinds of critical points may be similar.

Apart from strictly mechanical models (\ref{pendulum1}--\ref{pendulum2}), in our investigations on perpetual points we have also analyzed systems of different types. One of such examples is the polynomial system:
\begin{eqnarray}
   \left\{
     \begin{array}{l}
     \dot{x} = 9 x - 2 x^3 + 9 y - 2 y^3 - 1,\\
		 \dot{y} = -11 x + 2 x^3 + 11 y - 2 y^3 + 1,
     \end{array}
   \right.
	\label{quasiSys}
\end{eqnarray}
which has been introduced by Zhou et al. [2012]. They have studied the concept of quasi--potential landscape in complex multistable systems, and showed that using their method the quasi--potential function of (\ref{quasiSys}) can be described as $V_{x y} = -5 (x^2 + y^2) + 0.5 (x^4 + y^4) + xy + x$.

In Fig.~\ref{fig2}(c) the color plot of $V_{x y}$ has been presented for phase space $(x,y) \in {[-3,3]}^2$. System (\ref{quasiSys}) has 4 stable and 5 unstable fixed points, which are showed by white circles and black squares respectively, and are located in the extrema of quasi--potential. We have found 8 perpetual points in considered system (these are marked by gray stars). Using latter points, all the stable equilibria can be located. Gray arrows in Fig.~\ref{fig2}(c) denote which equilibrium can be reached from which point.

To examine the transitions between critical points of the system (\ref{quasiSys}), we have studied the trajectories starting from these points, which lead to one stable equilibrium, namely SFP1=$(2.3004, -2.3433)$ (red circle in Fig.~\ref{fig2}(d)). Considered point can be located from unstable fixed points UFP1=$(-0.1234, -2.2299)$, UFP2=$(0.1012, 0.0101)$ and UFP3=$(2.1724, 0.2194)$ (blue squares), as well as from perpetual points PP1=$(1.293, -2.6569)$ and PP2=$(0.7459, -1.2813)$ (green stars). Boundaries of region shown in Fig.~\ref{fig2}(d) are marked as the black inbox in Fig.~\ref{fig2}(c). As one can see, green curves corresponding to perpetual points converge to the blue trajectory starting from UFP1. Point PP1 is located close to it, and its orbit reaches the curve quickly, while trajectory beginning from PP2 needs more transient time. We have examined scenarios between other critical points of system (\ref{quasiSys}) and many similar transitions can be observed. 

As one can see, in systems with defined potential (like the ones presented above, eqs. (\ref{pendulum1}-\ref{quasiSys})) perpetual points have allowed us to localize all the stable equilibria present in the system. Although the latter points can be received explicitly from the equations of motion and there is no need to develop methods of finding them, obtained results extends our knowledge of potential applications of perpetual points. Since they lead to all SFPs, is it possible that they may guide also to other stable states co--existing in the system? Or, when stable equilibrium is replaced with more complex state, can they locate periodic or chaotic attractors? Our investigation on trajectories starting from perpetual points shows that they may be very similar to the ones denoted by unstable fixed points. In some sense, the former flows copy the latter ones, and hence, both types of points may have similar dynamical properties. This similarity can become essential in analysis of multistable systems. In the case of self--excited attractors, the localization problem can be solved using standard UFPs. However, if the system possesses hidden states (there are no UFPs, or they can not locate all the attractors), the localization role of unstable equilibria may be taken by perpetual points. Such hypothesis is additionally motivated by the fact that the concept of perpetual points bases on the theory of stationary points, since the former points are generalization (higher order) of the latter ones. Some of the examples showing that described property of perpetual points may be used in more complex models are shown below.
\vspace{\baselineskip}

As have been presented, perpetual points can be found in most elementary dynamical systems, in which potential can be described explicitly. However, usually the problem of localization of co--existing attractors arise in complex multistable systems, especially ones with hidden oscillations. In such cases, the analysis of existence of perpetual points and their potential usage in localizing different attractors becomes crucial.

In our research, we have studied chaotic flows, in which hidden oscillations can be observed. One of the earliest example of such systems can be found in the studies of Nose [1984] and Hoover [1985], i.e. 
\begin{eqnarray}
   \left\{
     \begin{array}{l}
     \dot{x} = y,\\
		 \dot{y} = -x - yz,\\
     \dot{z} = \alpha (y^2 - 1), 
		\end{array}
   \right.
	\label{NoseHooverSystem}
\end{eqnarray}
where $\alpha \neq 0$ is the parameter.

Motivated by system (\ref{NoseHooverSystem}), many researchers proposed similar artificial flows containing hidden attractors. In our analysis we have examined the system described by Wei [2011], which is defined by the following equations:
\begin{eqnarray}
   \left\{
     \begin{array}{l}
     \dot{x} = -y,\\
		 \dot{y} = x + z,\\
     \dot{z} = 2 y^2 + xz -0.35. 
		\end{array}
   \right.
	\label{WeiSystem}
\end{eqnarray}

Systems (\ref{NoseHooverSystem}) and (\ref{WeiSystem}) does not have fixed points, but exhibit hidden oscillations. Because the regions of instability (from which trajectories escape to infinity) can be observed in both models, the problem of choosing appropiate initial conditions leading to desired final state becomes non--trivial.

\begin{figure}
\includegraphics[scale=0.55]{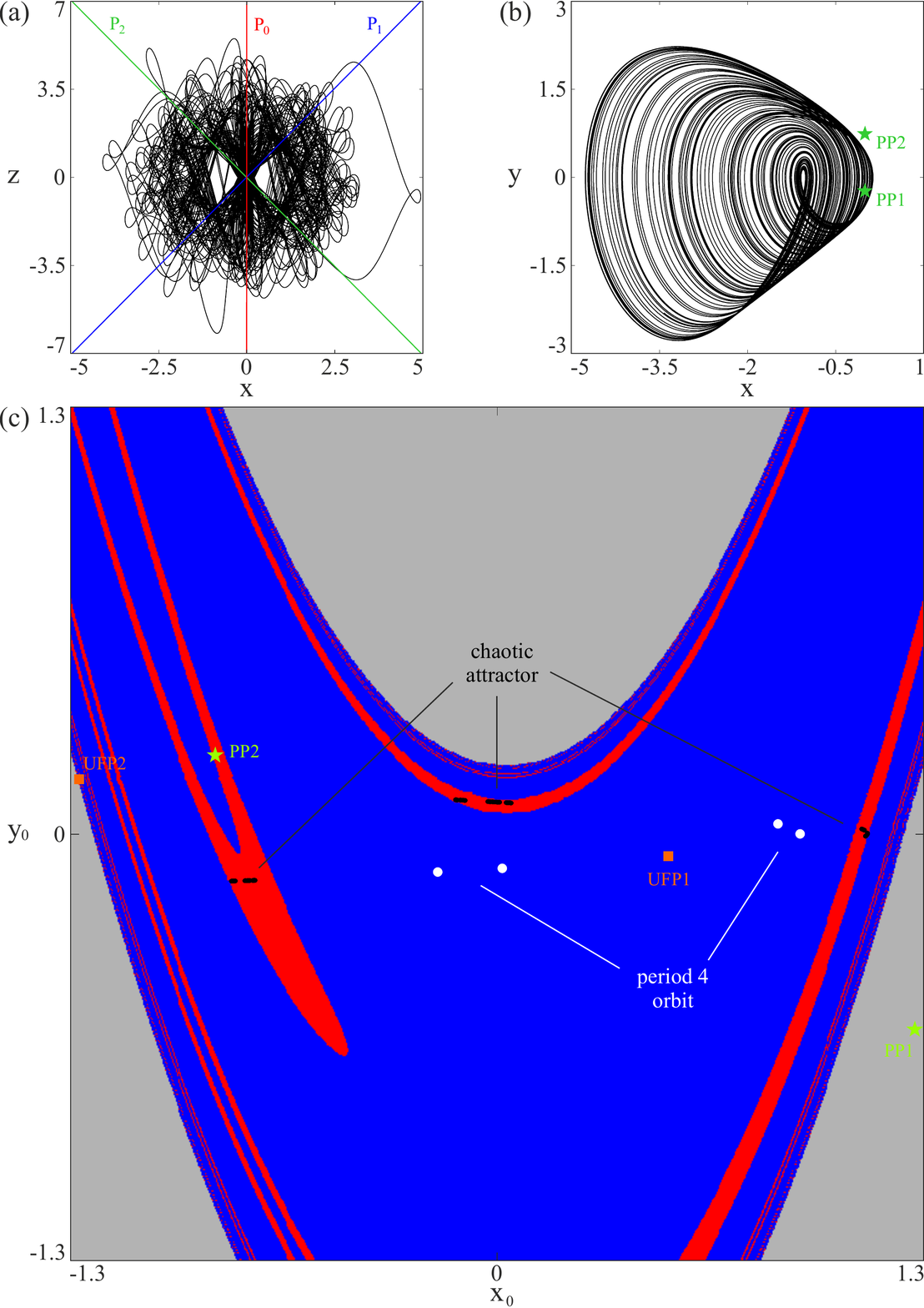}
\caption{(color online). In (a) the projection of chaotic trajectory (black) and perpetual points (sets $P_0$ (red), $P_1$ (blue), $P_2$ (green)) of system (\ref{NoseHooverSystem}) on $(x,z)$ plane is shown, while in (b) the projection of chaotic attractor (black) and perpetual points (green stars) of system (\ref{WeiSystem}) on $(x,y)$ plane are presented. The basins of attraction, co--existing attractors and critical points of map (\ref{henonMap}) are shown in (c).}
\label{fig3}
\end{figure}

In Fig.~\ref{fig3}(a--b) one can observe the results of our investigations on systems (\ref{NoseHooverSystem}, \ref{WeiSystem}). For any parameter value $\alpha$, system (\ref{NoseHooverSystem}) has infinitely many perpetual points, i.e. points from the line $P_0 = \left\{ (0,0,z^{*}): z^{*} \in \mathbb{R} \right\}$ are perpetual. Nonetheless, if $\alpha \in (-\infty, 0) \cup [1, \infty)$, there co--exist two additional curves of perpetual points, namely $P_1 = \left\{ (\sqrt{-1/\alpha + 1} z^{*}, - \sqrt{-1/\alpha + 1}, z^{*}): z^{*} \in \mathbb{R} \right\}$ and $P_2 = \left\{ (-\sqrt{-1/\alpha + 1} z^{*}, \sqrt{-1/\alpha + 1}, z^{*}): z^{*} \in \mathbb{R} \right\}$. In Fig.~\ref{fig3}(a) described sets $P_0, P_1, P_2$ for $z^{*} \in [-7, 7]$ and hidden oscillations that can be observed in system (\ref{NoseHooverSystem}) are shown as the projection on $(x, z)$ plane for parameter value $\alpha=2$. The curve $P_0$ is presented in red, while the lines $P_1$ and $P_2$ are marked in blue and green respectively. Part of the hidden chaotic trajectory is shown in black. In our calculations we have observed that points from $P_0$ set belong to instability region of the system, and consequently lead to infinity. However, using perpetual points from $P_1$ or $P_2$, one can obtain hidden attractor. Both curves $P_1, P_2$ have been examined for $z^{*} \in [-20, 20]$, and all obtained trajectories exhibit chaotic oscillations.

On the other hand, system (\ref{WeiSystem}) has two co--existing perpetual points, namely PP1 $= (0, (1-\sqrt{3.8})/4, 0) \approx (0, -0.2373, 0)$ and PP2 $= (0, (1+\sqrt{3.8})/4, 0) \approx (0, 0.7373, 0)$, which projection on $(x,y)$ plane is shown in Fig.~\ref{fig3}(b) as green stars. The chaotic attractor of (\ref{WeiSystem}) is presented in black. Considering calculated points as initial conditions, one can observe that the former point (PP1) leads to infinity, while the latter ones (PP2) locate chaotic state (attractor can be reached from it). Hence, in both systems (\ref{NoseHooverSystem}, \ref{WeiSystem}) perpetual points have been useful in localization of hidden attractors.
\vspace{\baselineskip}

Although the concept of perpetual points has been originally introduced in dynamical systems given by ordinary differential equations, it can be also generalized into models of different structure. Recently, this kind of critical points has been described in discrete--time systems, i.e. maps [Dudkowski et al., 2016a].

The idea of perpetual points in maps is based on the concept of \textit{discrete derivative} [Dudkowski et al., 2016a], which is an analogy to the standard derivative for continuous functions. If we consider the $n$--dimensional map given by equations:
\begin{equation}
x^{i}_{t+1} = f_{i} (x^{1}_{t}, \ldots, x^{n}_{t}),
\label{mapDD}
\end{equation}
where $i=1, \ldots, n$ and $t \in \mathbb{N} \cup \left\{ 0 \right\}$ is discrete time, the \textit{discrete derivative} of (\ref{mapDD}) will be defined as the following sequence:
\begin{equation}
d x^{i}_{t} := \frac{x^{i}_{t+1}-x^{i}_{t}}{t+1-t} = x^{i}_{t+1}-x^{i}_{t} = f_{i} (x^{1}_{t}, \ldots, x^{n}_{t}) - x^{i}_{t},
\label{d-der}
\end{equation}
where $i \in  \left\{ 1, \ldots, n \right\}$ is the chosen space variable. Since similar transformation can be applied to (\ref{d-der}), we may obtain sequence $d^{2} x^{i}$, referred as the \textit{second order discrete derivative} of system (\ref{mapDD}). According to defined relations, we we can introduce the concept of perpetual points in maps. Namely, we will call the point $(x^{1}_{pp}, \ldots, x^{n}_{pp})$ for which $d^{2} x^{i}_{pp} = 0 \wedge  d x^{i}_{pp} \neq 0, i=1, \ldots, n$ the \textit{perpetual point} of system (\ref{mapDD}).

The thorough analysis of perpetual points in maps, as well as their generalization in the form of \textit{perpetual loci} can be found in [Dudkowski et al., 2016a]. Below we present the application of introduced concept in simple example of bi--stable discrete--time system.

Let us consider the Henon map [D'Alessandro et al., 1990; Biham \& Wenzel, 1990; Jiang, 2016], given by the set of equations:
\begin{eqnarray}
   \left\{
     \begin{array}{l}
     x_{t+1}=1-a x^{2}_{t} + y_{t},\\
		 y_{t+1}=-b x_{t},
     \end{array}
   \right.
	\label{henonMap}
\end{eqnarray}
where $a$ and $b$ are the parameters.

In Fig.~\ref{fig3}(c) we have presented the results of our studies on system (\ref{henonMap}) for $a=1.5$ and $b=0.13$. Considered map has two co--existing attractors: chaotic (black circles) and period 4 orbit (white ones). The basins of attraction of these states are shown in red and blue respectively, while the instability region is marked in gray. System (\ref{henonMap}) has two unstable equilibria, namely UFP1 $= (0.5225, -0.0679)$ and UFP2 $= (-1.2759, 0.1659)$, which are denoted by orange squares. The former point lies in the blue area, and lead to periodic solution. On the other hand, point UFP2 is located very close to the border between blue/red and gray basins. We have examined initial conditions in the very close neighborhood of this fixed point, and the trajectories starting from them escape to infinity or get attracted by one of the co--existing states. However, the probability that the trajectory will converge to chaotic attractor is significantly lower in comparison with the probability that it will dissipate or stabilize on periodic orbit. It remains unclear what precision of point UFP2 should be considered to locate desired state. Hence, the chaotic attractor of model (\ref{henonMap}) can be termed as the hidden one. On the other hand, map (\ref{henonMap}) has two perpetual points, PP1 $= (1.2734, -0.5945)$ and PP2 $= (-0.8667, 0.2401)$ (both marked by green stars). The former point leads to infinity, while the latter ones lies in the red basin. Therefore, the hidden chaotic state can be located directly using perpetual point PP2. In [Dudkowski et al., 2016a] we have investigated more closely the concept of perpetual points in maps, in relation with both self--excited and hidden attractors.
\vspace{\baselineskip}

In systems presented above we have shown that perpetual points may be useful in locating hidden attractors. Usually, the problem of finding hidden states is based on choosing random initial conditions and tracing the trajectories starting from them. Although this method is the most natural, it is also very time--consuming (especially for the high dimensional systems) and its results are uncertain. The main problem related with this procedure is to choose the proper subspace of phase space, from which initial conditions will be drawn. For unknown system it is unclear what this subspace should be. Moreover, if it's chosen arbitrarily and does no intersect with the basin of hidden attractor, the desired state will not be located at all. This issue does not arise for tool such as perpetual points, because we obtain exact values of conditions to examine.

\begin{figure}
\includegraphics[scale=0.5]{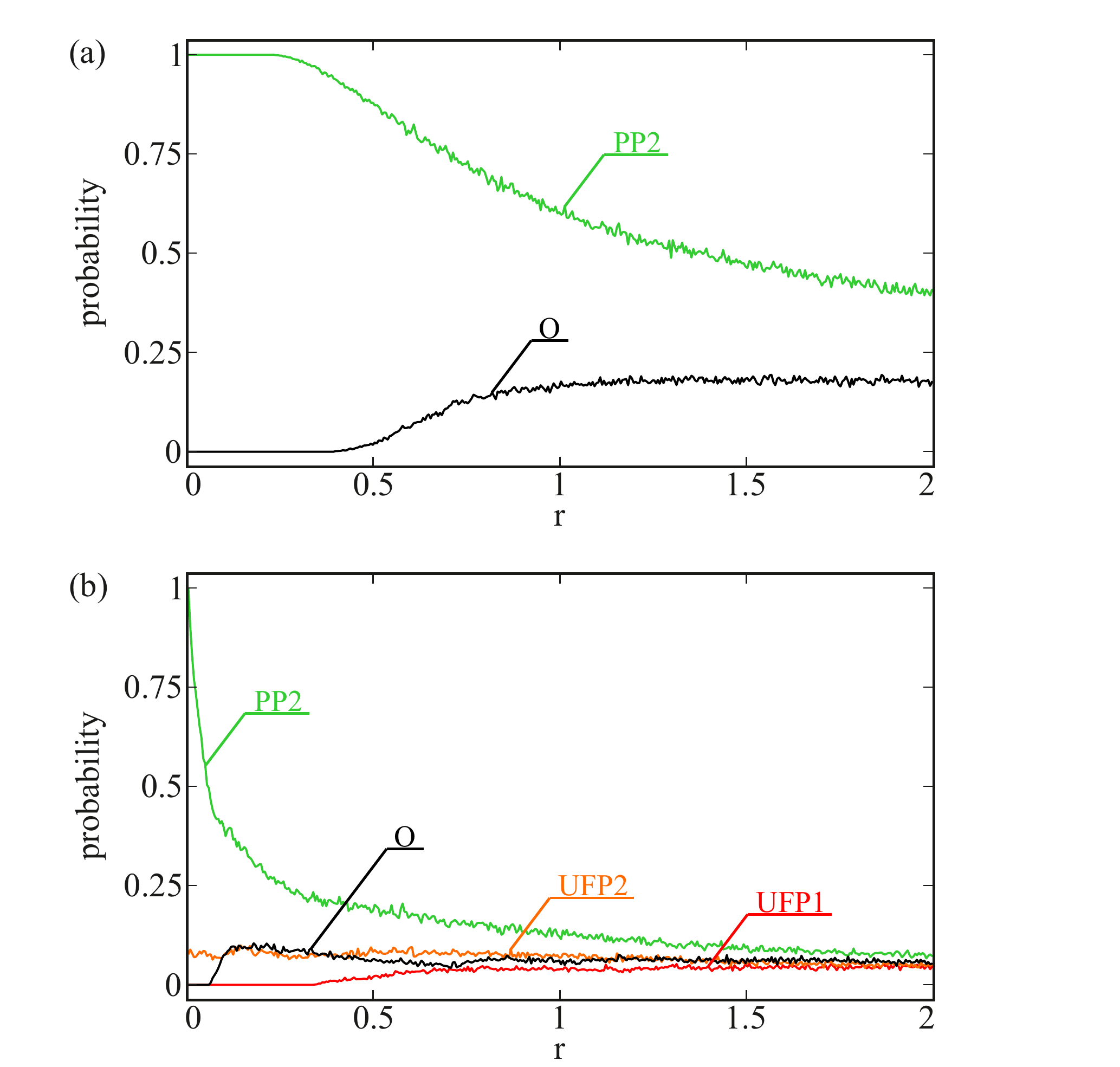}
\caption{(color online). Probability of reaching hidden attractors of systems (\ref{WeiSystem}) (a) and (\ref{henonMap}) (b), when initial conditions are drawn from a ball with a center in chosen critical point. Each curve corresponds to a different center point (marked in the Figure), and $r$ denotes the radius of considered ball.}
\label{fig4}
\end{figure}

To compare both methods, we have calculated the chances of reaching hidden attractors when using perpetual points and subspaces of random initial conditions. Probabilities have been estimated using typical Monte Carlo method. Our results are presented in Fig.~\ref{fig4}. Each curve corresponds to different center of the ball from which initial values are randomly chosen. The radius of such ball is denoted on the horizontal axis, while the probability of reaching hidden state is marked on the vertical one.
 
In Fig.~\ref{fig4}(a) results for system (\ref{WeiSystem}) are presented. Because the model does not have any fixed points, we have examined the phase space origin (denoted by O=$(0,0,0)$) and point PP2 (black and green curves respectively). As one can see, the probability of reaching hidden attractor of system (\ref{WeiSystem}) (black curve in Fig.~\ref{fig3}(b)) using perpetual point PP2 is guaranteed even if the initial conditions are chosen only in its close neighborhood. This means that PP2 is located in the dense region of basin of attraction. The probability begins to decrease, if we increase the radius to about $r>0.23$. On the other hand, if initial values are chosen from the ball with center O and radius $r<0.39$, all the trajectories will escape to infinity. The chance to observe chaotic attractor begins to increase slowly along with the increase of radius $r$. Nonetheless, the green curve remains above the black one for presented radius range. It should be noted that the whole hidden state is closed in the ball with radius $r=8.66$, and the probability of reaching it using random initial values drawn from such subspace equals 0.16. This example shows that inappropriate choice of initial conditions subspace may not only lower our chance to locate desired state, but even make it impossible.

On the other hand, the study of map (\ref{henonMap}) is shown in Fig.~\ref{fig4}(b). We have considered balls with centers placed in plane origin O=$(0,0)$ and three critical points: PP2, UFP1 and UFP2 (black, green, red and orange curves respectively). As one can see, the hidden chaotic attractor (shown as black dots in Fig.~\ref{fig3}(c)) can be reached from perpetual point PP2, but the probability dramatically decreases when increasing the radius of ball. In the case of origin O, chaotic state is first unavailable for $r<0.06$. Then, with increase of the radius the probability stabilizes at about 0.08. We can observe similar situation for UFP1, however in this case the minimum radius for crossing hidden attractor's basin equals $r \approx 0.34$. For the second stationary point UFP2, the locating potential seems to be independent from the $r$ value and oscillates around 0.07. The chance of finding chaotic attractor when initial conditions are drawn from the ball with $r=1.838$ (it contains whole subspace shown in Fig.~\ref{fig3}(c)) equals 0.06. As in the previous example, for a fixed radius $r$ the probability is the highest when ball is located in perpetual point. This shows that even though hidden oscillations can be traced using arbitrarily chosen initial conditions, it may be less effective and more time--consuming than considering perpetual points.
\vspace{\baselineskip}

The problem of finding co--existing attractors may also become non--trivial due to complicated structure of basins of attractors. An example of such structure is a \textit{riddled basin} [Chaudhuri \& Prasad, 2014; Alexander et al., 1992; Ott et al., 1994; Lai \& Grebogi, 1996], in which all points have pieces of another attractor's basin arbitrarily closely nearby [Ott et al., 1994]. In this case the usage of equilibria to locate all the states present in the system is not straightforward. If the value of fixed point is obtained numerically, the precision of chosen approximation may influence on the final state to which the point will lead. On the other hand, even if we know the value of equilibrium exactly, the choice of perturbation step needed to escape from it may result in different final states.

As an example, let us consider a complex quadratic map:
\begin{equation}
T(z_{t+1})=z^{2}_{t}-(1-\lambda i) \bar{z},
\label{Tmap}
\end{equation}
where $\lambda$ is the parameter fixed at $\lambda=1.02871376822$. System (\ref{Tmap}) has three co--existing chaotic attractors with riddled basins and its study can be found in [Alexander et al., 1992; Lopes, 1992; Schultz et al., 2017].

Map (\ref{Tmap}) possesses 4 fixed points, i.a. $z^{*} \approx -1.0287137682200000466 i$. Considering two approximations of equilibrium $z^{*}$, i.e. $z^{*}_{1}=-1.0287137682 i$ and $z^{*}_{2}=-1.028713768 i$ one can observe that trajectories starting from them will converge to two different attractors. Hence, the system behavior extremely depends on the chosen accuracy and it is impossible to predict the final state for unknown perturbation value. On the other hand, equation (\ref{Tmap}) has 5 perpetual points, which can be used to locate all the attractors co--existing in the system. Due to the fact that system is not in a stationary state in a perpetual point (in contrast to fixed point), its value can be chosen with arbitrarily high precision. Consequently, using perpetual points we know exactly to which state the system will converge.

\section{Conclusions}

In conclusion, we have shown that perpetual points can be widely observed in many different types of dynamical systems. They allowed us to localize all the fixed points co--existing in systems with defined potential, like the damped oscillators, or more artificial systems (e.g. polynomial model (\ref{quasiSys}) with quasi--potential). Trajectories starting from some of the perpetual points may be similar to the orbits running from unstable equilibria, which may induce that former points can localize attractors in the system. As have been presented, the concept of perpetual points may be also used for localization of co--existing attractors in more complex models (like artificial chaotic flows or maps). The described tool becomes even more essential due to development of research on hidden oscillations in dynamical systems. The study of latter states is crucial during the investigations of systems dynamics, and each new tool that allows to locate hidden attractors requires thorough analysis. As we have shown, perpetual points may become such a tool, and consequently their analysis in dynamical systems may be useful in finding hidden states (which is an analogy to localization of self--excited attractors by unstable fixed points). However, it should be emphasized that the concept has its natural limitations. Firstly, the model may be highly multistable, and consequently the number of co--existing attractors may exceed the number of perpetual points. Moreover, there is no guarantee that the system will possess any perpetual point (similarly to the possible absence of stationary points), while still having many attractors. All such limitations should be taken into account while considering potential application of perpetual points in multistable systems. Nevertheless, obtained results can be generalized into other systems with various types of dynamics, and consequently perpetual points may be used in analysis of co--existence problems in complex systems.
\vspace{\baselineskip}

\textbf{Acknowledgment}

This work has been supported by the National Science Centre, Poland, PRELUDIUM Programme -- Project No 2016/21/N/ST8/01877 and MAESTRO Programme -- Project No 2013/08/A/ST8/00/780. AP acknowledges the DST, Govt. of India for financial support, and thanks the TU, Lodz, for warm hospitality during several visits.
\vspace{\baselineskip}

\noindent
\textbf{References}
\vspace{\baselineskip}

\noindent
Alexander, J. C., Yorke, J. A., You, Z. \& Kan, I. [1992] "Riddled basins," \textit{Int. J. Bifurcation Chaos} \textbf{02}, 795-813.
\vspace{\baselineskip}

\noindent
Baker, G. L. \& Blackburn, J. A. [2005] \textit{The pendulum. A case Study in Physics} (Oxford University Press).
\vspace{\baselineskip}

\noindent
Barger, V. \& Olsson, M. G. [1973] \textit{Classical mechanics: a modern perspective} (McGraw--Hill).
\vspace{\baselineskip}

\noindent
Biham, O. \& Wenzel, W. [1990] "Unstable periodic orbits and the symbolic dynamics of the complex Henon map," \textit{Phys. Rev. A} \textbf{42}, 4639-4646.
\vspace{\baselineskip}

\noindent
Blekhman, I. [2003] \textit{Selected Topics in Vibrational Mechanics} (World Scientific Publishing Company).
\vspace{\baselineskip}

\noindent
Chaudhuri, U. \& Prasad, A. [2014] "Complicated basins and the phenomenon of amplitude death in coupled hidden attractors," \textit{Phys. Lett. A} \textbf{378}, 713-718.
\vspace{\baselineskip}

\noindent
D'Alessandro, G., Grassberger, P., Isola, S. \& Politi, A. [1990] "On the topology of the Henon map," \textit{J. Phys. A: Math. Gen.} \textbf{23}, 5285-5294.
\vspace{\baselineskip}

\noindent
Dudkowski, D., Jafari, S., Kapitaniak, T., Kuznetsov, N. V., Leonov, G. A. \& Prasad, A. [2016] "Hidden attractors in dynamical systems," \textit{Phys. Rep.} \textbf{637}, 1-50.
\vspace{\baselineskip}

\noindent
Dudkowski, D., Prasad, A. \& Kapitaniak, T. [2015] "Perpetual points and hidden attractors in dynamical systems," \textit{Phys. Lett. A} \textbf{379}, 2591-2596.
\vspace{\baselineskip}

\noindent
Dudkowski, D., Prasad, A. \& Kapitaniak, T. [2016a] "Perpetual points and periodic perpetual loci in maps," \textit{Chaos} \textbf{26}, 103103.
\vspace{\baselineskip}

\noindent
Hoover, W. G. [1985] "Canonical dynamics: Equilibrium phase-space distributions," \textit{Phys. Rev. A} \textbf{31}, 1695-1697.
\vspace{\baselineskip}

\noindent
Jafari, S., Nazarimehr, F., Sprott, J.C. \& Golpayegani, S. M. R. H. [2015] "Limitation of Perpetual Points for Confirming Conservation in Dynamical Systems," \textit{Int. J. Bifurcation Chaos} \textbf{25}, 1550182.
\vspace{\baselineskip}

\noindent
Jiang, H., Liu, Y., Wei, Z. \& Zhang, L. [2016] "Hidden chaotic attractors in a class of two-dimensional maps," \textit{Nonlinear Dynam.} \textbf{85}, 2719-2727.
\vspace{\baselineskip}

\noindent
Kibble, T. W. B. \& Berkshire, F. H. [2004] \textit{Classical Mechanics} (Imperial College Press).
\vspace{\baselineskip}

\noindent
Kovacic, I. \& Brennan, M. J. [2011] \textit{The Duffing Equation: Nonlinear Oscillators and their Behaviour} (Wiley).
\vspace{\baselineskip}

\noindent
Lai, Y.-C. \& Grebogi, C. [1996] "Characterizing riddled fractal sets," \textit{Phys. Rev. E} \textbf{53}, 1371-1374.
\vspace{\baselineskip}

\noindent
Leonov, G. A. \& Kuznetsov, N. V. [2013] "Hidden Attractors in Dynamical Systems: from Hidden Oscillations in Hilbert-Kolmogorov, Aizerman, and Kalman Problems to Hidden Chaotic Attractor in Chua Circuits," \textit{Int. J. Bifurcation Chaos} \textbf{23}, 1330002.
\vspace{\baselineskip}

\noindent
Leonov, G. A., Kuznetsov, N. V. \& Vagaitsev, V. I. [2011] "Localization of hidden Chua's attractors," \textit{Phys. Lett. A} \textbf{375}, 2230-2233.
\vspace{\baselineskip}

\noindent
Lopes, A. O. [1992] "On the Dynamics of Real Polynomials on the Plane," \textit{Comput. Graph.} \textbf{16}, 15-23.
\vspace{\baselineskip}

\noindent
Nazarimehr, F., Saedi, B., Jafari, S. \& Sprott, J. C. [2017] "Are perpetual points sufficient for locating hidden attractors?" \textit{Int. J. Bifurcation Chaos}.
\vspace{\baselineskip}

\noindent
Nose, S. [1984] "A molecular dynamics method for simulations in the canonical ensemble," \textit{Mol. Phys.} \textbf{52}, 255-268.
\vspace{\baselineskip}

\noindent
Ott, E. [1993] \textit{Chaos in Dynamical Systems} (Cambridge University Press).
\vspace{\baselineskip}

\noindent
Ott, E., Alexander, J. C., Kan, I., Sommerer, J. C. \& Yorke, J. A. [1994] "The transition to chaotic attractors with riddled basins," \textit{Physica D} \textbf{76}, 384-410.
\vspace{\baselineskip}

\noindent
Parlitz, U. \& Lauterborn, W. [1985] "Superstructure in the bifurcation set of the Duffing equation $\ddot{x} + d \dot{x} + x + x^3 = f \cos (\omega t)$," \textit{Phys. Lett. A} \textbf{107}, 351-355.
\vspace{\baselineskip}

\noindent
Pisarchik, A. N. \& Feudel, U. [2014] "Control of multistability," \textit{Phys. Rep.} \textbf{540}, 167-218.
\vspace{\baselineskip}

\noindent
Prasad, A. [2015] "Existence of Perpetual Points in Nonlinear Dynamical Systems and Its Applications," \textit{Int. J. Bifurcation Chaos}, \textbf{25}, 1530005.
\vspace{\baselineskip}

\noindent
Prasad A. [2016] "A Note On Topological Conjugacy For Perpetual Points," \textit{Int. J. Non. Sc.} \textbf{21}, 60-64.
\vspace{\baselineskip}

\noindent
Sabarathinam, S., Thamilmaran, K., Borkowski, L., Perlikowski, P., Brzeski, P., Stefanski, A. \& Kapitaniak, T. [2013] "Transient chaos in two coupled dissipatively perturbed Hamiltonian Duffing oscillators," \textit{Commun. Nonlinear Sci. Numer. Simul.} \textbf{18}, 3098-3107.
\vspace{\baselineskip}

\noindent
Schultz, P., Menck, P. J., Heitzig, J. \& Kurths, J. [2017] "Potentials and Limits to Basin Stability Estimation," \textit{New J. Phys.} \textbf{19}, 023005.

\noindent
Ueta, T., Ito, D. \& Aihara, K. [2015] "Can a Pseudo Periodic Orbit Avoid a Catastrophic Transition?," \textit{Int. J. Bifurcation Chaos} \textbf{25}, 1550185.
\vspace{\baselineskip}

\noindent
Wei, Z. [2011] "Dynamical behaviors of a chaotic system with no equilibria," \textit{Phys. Lett. A} \textbf{376}, 102-108.
\vspace{\baselineskip}

\noindent
Wei, Z., Moroz, I., Wang, Z., Sprott, J. C. \& Kapitaniak, T. [2016] "Dynamics at infinity, degenerate Hopf and zero--Hopf bifurcation for Kingni--Jafari system with hidden attractors," \textit{Int. J. Bifurcation Chaos} \textbf{26}, 1650125.
\vspace{\baselineskip}

\noindent
Wei, Z., Pham, V.-T., Kapitaniak, T. \& Wang, Z. [2016a] "Bifurcation analysis and circuit realization for multiple--delayed Wang--Chen system with hidden chaotic attractors" \textit{Nonlinear Dynam.} \textbf{85}, 1635-1650.
\vspace{\baselineskip}

\noindent
Wei, Z., Wang, R. \& Liu, A. [2014] "A new finding of the existence of hidden hyperchaotic attractors with no equilibria," \textit{Math. Comput. Simulat.} \textbf{100}, 13-23.
\vspace{\baselineskip}

\noindent
Wei, Z., Yu, P., Zhang, W. \& Yao, M. [2015] "Study of hidden attractors, multiple limit cycles from Hopf bifurcation and boundedness of motion in the generalized hyperchaotic Rabinovich system," \textit{Nonlinear Dynam.} \textbf{82}, 131-141.
\vspace{\baselineskip}

\noindent
Wei, Z. \& Zhang, W. [2014] "Hidden hyperchaotic attractors in a modified Lorenz--Stenflo system with only one stable equilibrium," \textit{Int. J. Bifurcation Chaos} \textbf{24}, 1450127.
\vspace{\baselineskip}

\noindent
Wei, Z., Zhang, W., Wang, Z. \& Yao, M. [2015b] "Hidden Attractors and Dynamical Behaviors in an Extended Rikitake System," \textit{Int. J. Bifurcation Chaos} \textbf{25}, 1550028.
\vspace{\baselineskip}

\noindent
Zhou, J. X., Aliyu, M. D. S., Aurell, E. \& Huang, S. [2012] "Quasi-potential landscape in complex multi-stable systems," \textit{J. R. Soc. Interface} \textbf{9}, 3539-3553.
\vspace{\baselineskip}

\end{document}